
\input phyzzx
\REF\OK{Y. Ohnuki and S. Kamefuchi, `Quantum Field Theory and Parastatistics'
(Tokyo: University of Tokyo Press/ Berlin: Springer, 1982)}
\REF\RTF{For example, N. Yu. Reshetikhin, L.A. Takhtadzhyan and L.D. Faddeev
\journal Leningrad Math. J. &1 (90) 193.}
\REF\PW{W. Pusz and S.L. Woronowicz \journal Rep. Math. Phys. &27 (89) 231.}
\REF\PU{W. Pusz \journal Rep. Math. Phys. &27 (89) 349.}
\REF\KO{K. Odaka \journal J. Phys. A: Math. Gen. &25 (92) L39.}
\REF\B{J. Wess and B. Zumino \journal Nucl. Phys. B (Proc. Suppl.) &18B (90)
302.}
{\hsize=17.5truecm \leftskip=9.5cm
{NDA-FP-9/92}\par
\vskip -4mm
{October 1992}\par}

\title{Comment on $q$-deformation in Second Quantization Procedure}

\author{Kazuhiko Odaka}

\address{Department of Mathematics and Physics}
\address{National Defense Academy}
\address{Hashirimizu, Yokosuka 239, JAPAN}

\abstract{When the $q$-deformed creation and annihilation operators
are used in a second quantization procedure, the algebra satisfied by
basis vectors (orthogonal complete set) should be also deformed
such as a field operator remains invariant under the coaction of the quantum
group. In the 1+1 dimensional quantum field theories
we deform the algebra of the basis vectors and study the $q$-deformation
in the second quantization procedure. }

\endpage


In the quantum field theories a field operator is expanded in terms of an
orthogonal complete set which forms a set of basis vectors.
Each coefficient is identified with annihilation or
creation operator.  The field operator can be also expanded by other
orthogonal complete set and we can define other annihilation and creation
operators.  The two sets of basis vectors are related through linear
transformations
and then the operators transform linearly.
\par
In the quantum field theories, statistics follow as a consequence of field
quantization and then statistics must be independent of the choice of the
orthogonal complete sets.  Thus, the same algebra as the operators
shall hold also for the operators transformed.  In other words, the algebra
satisfied by the operators must be covariant under the transformations.
 Such transformations are called canonical ones and they form a Lie group
in the standard cases.\refmark{\OK}
\par
Now, a class of non-commutative and non-cocommutative Hopf algebra has been
found in the investigations of the integral systems.  These Hopf algeras
are $q$-deformed function algebra of the Lie groups
 and they are generated by matrix elements
which do not commute each other.
 This structure is called quantum group.\refmark{\RTF}
\par
Recently many authors study the canonical transformations from the
viewpoint of the quantum group. In these studies \refmark{\PW - \KO}the
fermionic or bosonic
algebra of creation and annihilation operators, the so-called
oscillator algebra, is deformed to be covariant under the coaction of
the quantum group. If we identify the matrix elements appearing
 in the canonical transformations
with the generators of the quantum group and we apply the $q$-deformed
oscillator algebra to the quantum field, we should also deform
 the algebra of the basis vectors such as
the quantum field remains invariant under the coaction of the quantum group.
However, none has studied the deformation of the algebra
of the basis vectors.
\par
The purpose of this note is to investigate the
$q$-deformation in the second quantization procedure
in consideration of the $q$-deformation of the basis vectors.
\par

First, the standard case is summarized.
 We consider a single-component fermionic field $\psi(x,t)$ in the 1+1
dimensional space-time. Its canonical anti-commutation relations are
$$\eqalignno{\psi(x,t) \psi(y,t) + \psi(y,t) \psi(x,t) &= 0,\cr
\psi(x,t) \psi^\dagger (y,t) + \psi^\dagger (y,t) \psi(x,t)
&= \delta(x-y).&(1)\cr}$$
The field can be regarded as a non-relativistic de-Broglie one or as
a relativistic chiral one but, for simplicity, we suppose the first case.
The field operator $\psi (x,t)$ may be written as
$$\psi(x,t) = \sum_k f_k(x,t)a_k, \eqno(2)$$
where $f_k(x,t)$ 's  are some basis functions and
 the coefficients $a_k$ are identified with annihilation operators
whose algebra is
$$a_ka_p + a_pa_k = 0,~~~~~  a_ka^\dagger _p + a^\dagger _pa_k = \delta_{pk}.
\eqno(3)
          $$
\par
In the quantum field theories we need a cut-off procedure in the actual
calculations.
Thus we introduce the cut-off in the indices $k$ and define
a new field operator $\hat \psi (x,t)$ as
$$\hat \psi (x,t) = \sum_{N<k<M} f_k (x,t) a_k. \eqno(4)$$
The canonical anti-commutation relations of $\hat \psi (x,t)$ read
$$\eqalignno{\hat\psi (x,t) \hat\psi (y,t) + \hat\psi (y,t) \hat\psi (x,t)
&= 0,\cr
\hat\psi (x,t) \hat\psi^\dagger(y,t) + \hat\psi^\dagger (y,t) \hat \psi (x,t)
&= \sum_{N<k<M} f_k (y,t)f_k^\ast (x,t). &(5)\cr}$$
The quantum field $\hat \psi (x,t)$ may be also written as
$$\hat\psi (x,t) = \sum_{N<k<M} g_k(x,t) a'_k \eqno(6) $$
where
$$g_k(x,t) = \sum_{N<k<M} f_p(x,t)t^\dagger_{pk} ~~~ {\rm with} ~~
\sum _{N<k<M} t_{pk} t^\dagger_{kl} = \sum_{N<k<M} t^\dagger_{pk} t_{kl} = 1.
\eqno(7)$$
Since $f_k$ 's and $g_k$ 's are related by the unitary matrix, so are the
operators
$a_k$ 's and $a'_k$ 's (transformation theory):
$$a'_k = \sum_{N<p<M} t_{kp}a_p,~~~~~a'^\dagger_k = \sum_{N<p<M} a^\dagger_p
 (t^\dagger)_{pk}. \eqno(8)$$
\par
The oscillator algebra (3) is covariant under (8) and then even the quantum
field theories with the cut-off are formulated in such a way that the
consequences do not
depend on the choice of the basis functions.
When one removes the cut-off, the field operator $\hat \psi (x,t)$ becomes
the ordinary one $\psi (x,t)$ and the canonical anti-commutation
relations (5) reduce to the usual ones (1).
\par
Next, let us consider the case that the canonical transformations form a
quantum
group.  For simplicity we take two degrees of freedom and then the quantum
group $U_q(2)$ is realized by the canonical transformations.
Since the $q$-determinant of $U_q(2)$ is center, we may take this to be unity
and the canonical transformation matrix reads the quantum group $SU_q(2)$
defined by
$$\parallel T^i_j \parallel~=~\left( \matrix{a~~&b\cr
                                      c~~&d\cr}\right), $$
$$\parallel T^{\dagger i}_j \parallel~ = ~\left( \matrix{a^\ast~~&c^\ast\cr
b^\ast~~&d^\ast\cr}\right)
{}~=~\left( \matrix{d~~&-q^{-1}b\cr
                 -qc~~&a\cr}\right),\eqno(9)$$
where
$$ab = qba,~~bc =cb,~~ac = qca,~~bd =qdb,~~cd =qdc,$$
$$ad-da = (q-q^{-1})bc,~~det_qT \equiv ad-qbc =1.$$
Here, $q$ is a real number and this matrix satisfies the unitary condition,
$$T^\dagger T = T T^\dagger = 1. \eqno(10)$$
\par
We introduce a pair of fermionic annihilation operators $(A_1, A_2)$
transforming linearly under
$SU_q(2)$ :
$$A'_i = \sum_{j=1}^2 T_{ij} A_j,~~~
A'^\dagger_i = \sum_{j=1}^2 A_j^\dagger (T^\dagger)_{ji} \eqno(11)$$
where the operators $A_i$ 's and $A_i^\dagger$ 's commute with $T_{ij}$
and $(T^\dagger)_{ij}$.
The covariant algebra of $A_i$ 's and $A_i^\dagger$ 's is not unique and so
 conforming to ref.~4,  we use the algebra:
$$A_1^2 = 0,~~A_2^2 = 0,~~~A_1A_2 + q^{-1}A_2A_1 =0,$$
$$A_1^\dagger A_2 +q^{-1}A_2 A_1^\dagger = 0,\eqno(12)$$
$$A_1 A_1^\dagger + A_1^\dagger A_1 + (1 - q^2 ) A_2^\dagger A_2 = 1,~~~
A_2 A_2^\dagger + A_2^\dagger A_2 =1. $$
\par
The  quantum field $\hat\Psi (x,t)$, whose canonical relations will be
consistently
determined later, is decomposed such as
$$\hat\Psi (x,t) = \sum_{k=1}^2 F_k(x,t) A_k,\eqno(13)  $$
where $F_k(x,t)$ is a basis function.  We take another set of basis functions
 $G_k(x,t)$ and the quantum field may be rewritten as
$$\hat \Psi (x,t) = \sum_{k=1}^2 G_k(x,t) A'_k. \eqno(14)  $$
Since the operators $A_i$ 's transform in the manner (11), the two sets of
the basis functions should be related
through
$$G_k(x,t) = \sum_{i=1}^2 F_i(x,t) (T^\dagger)_{ik} \eqno(15) $$
 where $F_i$ 's are assumed to commute with $T^\dagger_{ij}$.
The algebra of $F_i$ 's must be covariant under (15) but
 it can not be uniquely determined from the covariance requirement.
Thus, in a manner analogous to the $q$-bosonic oscillator algebra given
in ref.~3
 we adopt the following algebra,
$$F_2(x,t) F_1(x,t) - q F_1(x,t)F_2(x,t) = 0,~~~
F_1^\ast(x,t) F_2(x,t) - q F_2(x,t) F_1^\ast(x,t) =0,$$
$$F_1^\ast(x,t) F_1(x,t) - q^2
F_1(x,t) F_1^\ast(x,t) = 0,\eqno(16)$$
$$F_2^\ast(x,t) F_2(x,t) -q^2 F_2(x,t) F_2^\ast(x,t) + (1 - q^2)F_1(x,t)
F_1^\ast(x,t) =0. $$
This can be regarded as the $q$-deformation of the algebra satisfied by
complex numbers.
\par
Our problem is to determine the algebras between the operators and the
basis functions and between the functions at different points.
In so doing we impose the following conditions:
\par
(i)~~These algebras are covariant under $SU_q(2)$.
\par
(ii)~The canonical relations satisfied by the quantum fields
 $\hat \Psi (x,t)$ and $\hat\Psi^\dagger (x,t)$ \par
{\leftskip 7truemm have the closed form.
\par}
\noindent
The algebras between the operators and the basis functions are obtained as
$$A_1 F_1(x,t) - f F_1(x,t) A_1 =0,~~~
A_2 F_1(x,t) -f q^{-1}F_1(x,t)A_2 =0,$$
$$A_1F_2(x,t) -f q^{-1} F_2(x,t) A_1 =0, \eqno(17)$$
$$A_2 F_2(x,t) -f F_2(x,t) A_2 + (q^{-2}-1)A_1 F_1(x,t)=0$$
and
$$A_1^\dagger F_1(x,t) - g q^{-1}F_1(x,t) A_1^\dagger =0,~~~
A_2^\dagger F_2(x,t) - g q^{-1}F_2(x,t) A_2^\dagger =0,$$
$$A_2^\dagger F_1(x,t) - g F_1(x,t) A_2^\dagger =0 \eqno(18)$$
$$A_1^\dagger F_2(x,t) - g F_2(x,t) A_1^\dagger
+ (q -q^{-1}) A_2^\dagger F_1(x,t) =0,$$
where $f$ and $g$ are real numbers.
 For the algebra between  $F$ and $F^\ast$ we have
$$F_1^\ast(x,t) F_1(y,t) - q^2 F_1(y,t)F_1^\ast(x,t) =0,$$
$$F_2^\ast(x,t) F_1(y,t) - q F_1(y,t) F_2^\ast(x,t) =0, \eqno(19)$$
$$F_2^\ast(x,t) F_2(y,t) -q^2 F_2(y,t) F_2^\ast(x,t)
+ (1 - q^2) F_1(y,t)F_1^\ast(x,t) =0.$$
For the algebra between  $F$ 's at different points one obtains
$$F_1(x,t) F_1(y,t) - q^{-2} F_1(y,t) F_1(x,t) =0,$$
$$ F_2(x,t) F_2(y,t) -  q^{-2} F_2(y,t) F_2(x,t) =0,$$
$$F_2(x,t) F_1(y,t) - q^{-1} F_1(y,t) F_2(x,t) =0, \eqno(20)$$
$$F_1(x,t) F_2(y,t) - q^{-1}  F_2(y,t) F_1(x,t)
+ (q -q^{-1})F_2(x,t) F_1(y,t) =0$$
where $x > y$.
\par
By using the above relations we can check that the operators
$\hat \Psi (x,t)$ and $\hat \Psi ^\dagger (x,t)$ satisfy the following
closed relations:
$$\hat \Psi (x,t) \hat \Psi (y,t) + \hat \Psi (y,t) \hat \Psi (x,t) = 0,$$
$$\hat \Psi (x,t) \hat \Psi^\dagger(y,t) + g^{-2} \hat \Psi^\dagger(y,t)
\hat \Psi (x,t) = \sum_{k=1}^2 F_k(x,t)F_k^\ast (y,t). \eqno(21)$$
If we take $f$ to be unity, the canonical relations of $\hat \Psi$ 's
and $\hat
\Psi ^\dagger$ 's have the same form as ones of $\hat \psi$ 's
and $\hat \psi ^\dagger$ 's.
\par
The results derived in this note are based on the special properties
of $SU_q(2)$.
Thus, in order to deal with many degrees of freedom, we need to change
a little the algebras (12) and (16) so as to apply the R-matrix formulation
of the quantum
group to our argument\refmark.{\KO} The $q$-deformation of
the basis functions in the general case  is under investigation and these
results
will be published elsewhere.
\par
An interesting question is about wavefunctions of $q$-deformed systems.
In the expressions (13) and (14), $F_k(x,t)$ 's and $G_k(x,t)$ 's
may be regarded as wavefunctions, and these are transformed to each other
through
matrices $T$.  Thus, we have a question whether these wavefunctions are
considered as the ones
of the quantum mechanics on the quantum plane\refmark{\B} in the same way
as the usual
relation between qunatum mechanics and field theories or not.
A proper study of this question is needed.
\ack{The author is deeply indebted to S. Kamefuchi
for stimulating discussions.  He would like to thank T.W.B. Kibble
and D.I. Olive for their warm hospitality at Imperial
College of Science, Technology and Medicine, where this work was completed}
\refout
\bye